# Astronomical Orientations of Bora Ceremonial Grounds in Southeast Australia

Robert S. Fuller[1,2], Duane W. Hamacher[1,3] and Ray P. Norris[1,2,4]

[1] Warawara - Department of Indigenous Studies, Macquarie University, NSW, 2109, Australia
Email: *robert.fuller1@students.mq.edu.au*

[2] Research Centre for Astronomy, Astrophysics & Astrophotonics, Macquarie University, NSW, 2109, Australia

[3] Nura Gili Centre for Indigenous Programs, University of New South Wales, NSW, 2052, Australia
Email: *d.hamacher@unsw.edu.au*

[4] CSIRO Astronomy and Space Science, PO Box 76, Epping, NSW, 1710, Australia
Email: *raypnorris@gmail.com*

**Abstract**

Ethnographic evidence indicates that bora (initiation) ceremonial sites in southeast Australia, which typically comprise a pair of circles connected by a pathway, are symbolically reflected in the Milky Way as the 'Sky Bora'. This evidence also indicates that the position of the Sky Bora signifies the time of the year when initiation ceremonies are held. We use archaeological data to test the hypothesis that southeast Australian bora grounds have a preferred orientation to the position of the Milky Way in the night sky in August, when the plane of the galaxy from Crux to Sagittarius is roughly vertical in the evening sky to the south-southwest. We accomplish this by measuring the orientations of 68 bora grounds using a combination of data from the archaeological literature and site cards in the New South Wales Aboriginal Heritage Information Management System database. We find that bora grounds have a preferred orientation to the south and southwest, consistent with the Sky Bora hypothesis. Monte Carlo statistics show that these preferences were not the result of chance alignments, but were deliberate.

**Notice to Aboriginal and Torres Strait Islander Readers**

This paper discusses bora ceremonies and contains the names of people who have passed away. The exact locations of these sites are concealed in order to protect them. The coordinates provided are within 10 km of the site and are used only to demonstrate the general distribution of the known bora grounds in southeast Australia.

**Keywords:** Aboriginal Australian, Cultural Astronomy, Ceremonial Sites

## Introduction

In traditional Aboriginal cultures across Australia, young males are taught the laws, customs and traditions of the community and undergo a transition ceremony from boyhood to manhood. This ceremony often includes a 'rite of passage' event in which the initiated males undergo some form of body modification (Jacob 1991), typically involving tooth evulsion in southeast Australia (e.g. Berndt 1974:27–30). This ceremony goes by many names, but in Queensland (Qld) and New South Wales (NSW) it has come to be generally known as 'bora', the name used by the Kamilaroi of north-central NSW (Ridley 1873:269). Bora grounds generally consist of two circles of differing diameter connected by a pathway. The larger circle is regarded as a public space, while the smaller circle some distance away is restricted to initiates and elders. Bora ceremonies were one of the first Aboriginal cultural activities described by early Australian colonists in the Sydney region (Collins 1798:468-480; Hunter 1793:499-500). Because information about bora ceremonies is culturally sensitive, here we limit discussion of the ceremony itself.

There is a variety of evidence from the anthropological literature (e.g. Berndt 1974; Love 1988; Winterbotham 1957) that bora ceremonies are related to the Milky Way and that ceremonial grounds are oriented to the position of the Milky Way in the night sky at particular times of the year.

In this paper, we begin by exploring connections between bora ceremonies and the Milky Way using ethnographic and ethnohistoric literature. We then use the archaeological record to determine if bora grounds are oriented to the position of the Milky Way at particular times of the year. Finally, we use Monte Carlo statistics to see if these orientations were deliberate or the result of chance.

**Bora Ceremonial Grounds**

The layout of bora grounds is similar across southeast Australia, with only minor differences from region to region (Bowdler 2001:3; Mathews 1894:99). Several reports (e.g. Black 1944; Collins 1798:391; Fraser 1883; Howitt 1904; Mathews 1897a) describe the grounds as consisting of two rings of different sizes, connected by a pathway. In some places, Bora sites may comprise three or more rings (Steele 1984; Bowdler 2001). The border of each ring is made of raised earth or stone, and the area within is cleared of debris and the earth is stamped until firm. The larger ring, which is considered public, has a typical diameter of 20–30 m. The smaller ring (generally 10–15 m diameter) is considered the sacred area, where body modification takes place, and is restricted to initiates and elders. The two rings are connected by a pathway that ranges from a few tens to a few hundred metres in length. In 2004, an Aboriginal man from Marulan, NSW advised that parts of many such sites were destroyed immediately after the ceremony to conceal their location (Hardie 2004). For this reason, some bora sites reported in the archaeological literature feature only a single circle, with the smaller, sacred circle having been destroyed.

Bora grounds are distributed throughout southeast Australia, covering most of NSW and southern Qld and may extend into South Australia (Howitt 1904:501–508) and northern Qld (Roth 1909). Ceremonial rings, which may be bora grounds, have been found near Sunbury, Victoria (Vic.), although there are no ethnographic records attesting to their ceremonial use (Frankel 1982). Howitt (1904:512) cited a western boundary for bora running from the mouth of the Murray River to the Gulf of Carpentaria. Mathews (1897b:114) noted the bora can be found across three-quarters of NSW and some distance into western Qld, with a boundary extending from Twofold Bay near Eden, NSW in the south, to Moulamein, NSW in the west, and Barringun, Qld in the north. We recognise that the geographical area covered by this paper includes several distinct language groups, each of whom may have separate culture and traditions, and it may be misleading to aggregate the data from such a wide area. However, (a) the existence of similarly constructed bora rings implies some commonality in culture, and (b) aggregating orientations from a large geographic area will dilute any preferred orientations arising from a single Aboriginal group, rather than forming a correlation of spurious significance.

According to Love (1988), the bora ceremonies were predominantly held in August each year, although other authors report a variety of dates including March–May (Winterbotham 1957), April–June (Mathews 1894:99), May–July (Mathews 1894), August (Needham 1981:70), September–November (Winterbotham 1957), and October–December (Mathews 1894). This suggests that, in some cases, a number of variables influence the date of the bora ceremony, including the availability of food and water or having a sufficient number of boys to initiate (e.g. Mathews 1910). While these factors vary across the region, here we test the hypothesis advanced by Love (1988) who presents evidence associating the bora ceremony with the night sky and the orientation of the Milky Way, and suggests that most initiation ceremonies occur around August.

**Anthropological Support of an Astronomical Connection**

It is well established that the night sky plays a significant role in several Aboriginal cultures (e.g. Cairns and Harney 2004; Johnson 1998; Norris and Hamacher 2009; Hamacher 2012). In Aboriginal astronomical traditions, dark spaces within the Milky Way are as significant as bright objects. Two animals symbolically link bora ceremonies to these dark spaces. One was a spiritual serpent, commonly referred to as the 'Rainbow Serpent' across Australia, traced out by the curving dust lanes in the Milky Way. Needham (1981:69) explained that in Aboriginal communities of the Hunter Valley, motifs of the spiritual serpent were represented in bora ceremonies, with information about the serpent being recounted during the ceremony itself. The other animal was an emu, which is also traced by dust lanes in the Milky Way (Norris and Norris 2009). Love (1988:129–138) argued that the emu was an important part of the bora ceremony in southeast Australia, as did Berndt (1974:27–30), since male emus brood and hatch the emu chicks and rear the young (Love 1987). This is symbolic of the initiation of adolescent boys by their male elders.

Needham (1981) provided an illustration of the night sky and associated stars in local Aboriginal astronomical traditions. The illustration, which cites the 'All Father' as the star Altair, provides the positions of celestial objects in August, 'the month when Aboriginal initiation ceremonies were held' (Needham 1981:70). During the early part of the night in August, the Milky Way stretches across the sky from the northeast to the southwest. Many early colonial reports referred to an Aboriginal religion based on a deity variously described as Baiame, Bunjil or Mungan-ngaua (Henderson 1832:147; Howitt 1904:490–491; Ridley 1873:268). These names roughly translate to 'father' or 'father of all of us' (Howitt 1904:491). According to Fraser (1883:208) and Howitt (1884:458), Baiame gave his son, Daramulan, to the people and it is through Daramulan that Baiame sees all. Baiame is worshipped at the bora ceremony (Ridley 1873:269) and Daramulan is believed to come back to the earth by a pathway from the sky (Fraser 1883:212). Eliade (1996:41) reported that Baiame 'dwells in the sky, beside a great stream of water' (i.e. the Milky Way) and various reports (e.g. Berndt 1974; Hartland 1898; Howitt 1884) have claimed that the wife of Baiame (or in some cases Daramulan) is an emu. Reports of Baiame, Daramulan and Bunjil come from various cultures across southeast Australia, resulting in variations in these reports. However, they share some features, such as a close connection between bora ceremonies and the Milky Way.

**Testing the Hypothesis**

To focus the discussion, we concentrate specifically on the hypothesis advanced by Love (1987, 1988), who argued that ancestral spirits in the heavens held bora ceremonies in the Milky Way, which we refer to as the 'Sky Bora'. Love based his work, in part, on Winterbotham (1957), who obtained information from a Jinibara man from southeast Qld named Gaiarbau. According to Winterbotham (1957:38), bora circles,

> ... were always oriented towards points of the compass, the larger one to the north, and the smaller to the south ... They conformed in this rule to the position of two dark (black) spaces (circles)—the Coal Sacks in the heavens.

According to Love (1988:130–131), the Jinibara account identifies the Sky Bora with the Emu in the Sky (Gaiarbau et al. 1982:77; Winterbotham 1957:46). The 'Coal Sacks', or *Mimburi*, to which Winterbotham referred, are a dark absorption nebula bordering the western constellations Crux (Southern Cross), Centaurus and Musca, representing the head of the emu, with the star BZ Crucis representing the eye. The dust lanes that run through the stars Alpha and Beta Centauri represent the neck, while the Galactic bulge, near the intersection of Sagittarius, Scorpius and Ophiuchus, represents the body. This area is the centre of our Milky Way galaxy. The dust lanes along the Milky Way through Sagittarius trace out the legs (Figure 1). The motif of the celestial emu is found across Australia (e.g. Cairns and Harney 2004; Norris and Hamacher 2009:13; Stanbridge 1861:302; Wellard 1983:51).

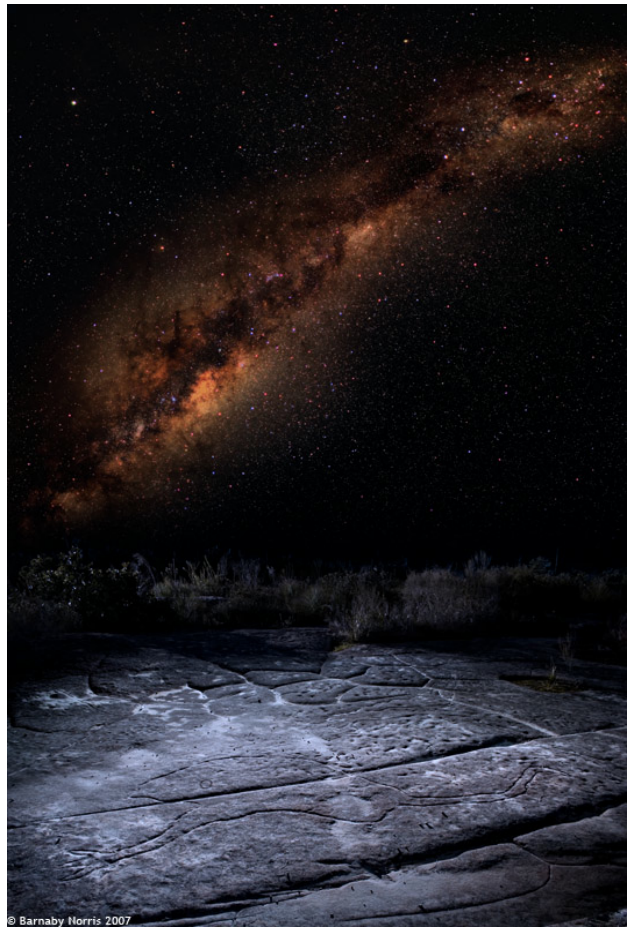

*Figure 1:* *The 'Emu in the Sky' visible over an emu engraving at Elvina Track in Kuringai Chase National Park north of Sydney, which may represent her celestial counterpart (image reproduced courtesy of Barnaby Norris).*

According to Winterbotham, other Aboriginal groups also knew of the dark nebulae, including the Badjala people of Fraser Island and the adjacent mainland, who call them *Wurubilum*, and the Wakka Wakka people near Murgon, Qld. This concept extends beyond southeast Australia. For example, Smith (1913) explained that during an initiation ceremony in Western Australia (WA) the initiate is left tied to the ground until the Milky Way is visible. He is then asked if 'he can see the two dark spots' and when he is able to see them, he is released. While this account is not from the area of this study, it may be similar to the example that Gaiarbau described.

Gaiarbau said that bora ceremonies were not held until the celestial bora rings returned to their 'proper points of the compass' (Winterbotham 1957:38). In clear winter skies in southeast Australia, the Milky Way is visible about an hour after sunset. The orientation of the plane of the Milky Way, as seen from southeast Australia an hour after sunset, changes from near vertical in the south-southeast in March to horizontal across the southern sky from east-southeast to west in June and back to vertical (but inverted) in the southwest in September. The Galactic bulge and the Coalsack (celestial emu) are not visible in the sky together an hour after sunset until May. At this time, the emu is not vertical in the sky (perpendicular to the horizon), but stretches from south to east.

The only time that the Sky Bora is vertically aligned to the horizon and can be seen in the sky together an hour after sunset is in August (or later in the night as the year progresses). The Galactic plane, which goes straight through the celestial emu, is vertical in August an hour or two after sunset (earlier in the evening later in the month, Figure 2). At this time, the azimuth is approximately 213º (south-southwest).

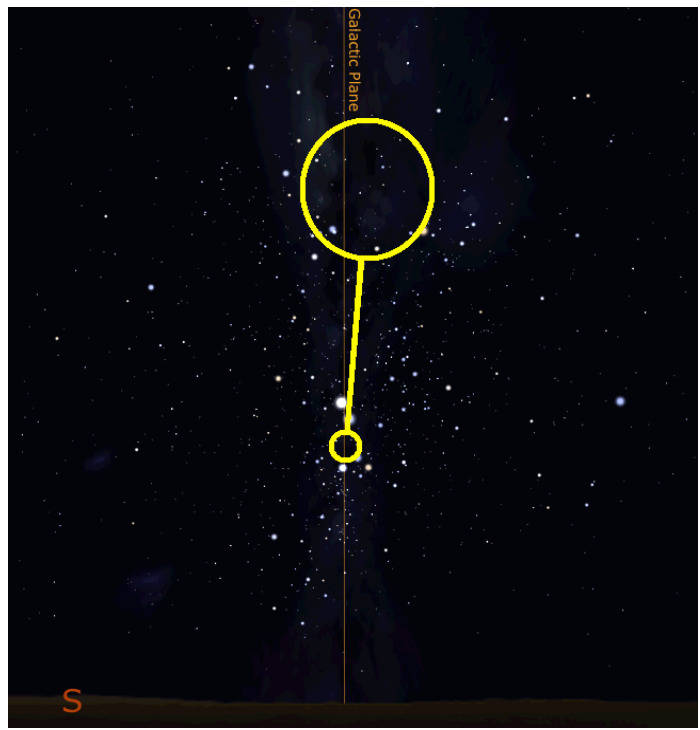

*Figure 2:* The sky bora in the Milky Way oriented vertically in the south-southwest sky in mid-August an hour after sunset, as seen from Brisbane. The large circle represents the larger bora circle and the body of the emu (near zenith). The smaller circle represents the Coalsack and the head of the emu. Graphics are from the Stellarium astronomical software package.

If Love's hypothesis is correct, we expect the orientation of each bora site from the larger circle to the smaller circle to be oriented to roughly 213º, corresponding to the time ceremonies are held (August). This expectation agrees with Needham (1981:70), who claimed that bora ceremonies in the Hunter Valley (NSW) were held in August when the Milky Way was vertical in the south-southwest.

Other researchers report that bora ceremonies in Qld and NSW are held at various times of the year, as noted in the previous section (Winterbotham 1957; Mathews 1894:99). Bora ceremonies may be held at times of the year that have little or nothing to do with the position of the Milky Way in the sky, even if the ceremonies are symbolically linked to the Milky Way. For example, Mathews (1894:128) claimed that the direction of one bora ring to the other 'is entirely dependent on the conformation of the country within which the ceremony is being held'. If bora grounds are not oriented to any particular object or direction, we expect to find a roughly uniform distribution in their orientations. However, if at least some bora grounds are oriented to the position of the Sky Bora, then we expect to find a preference for south-southwest orientations when we look at the overall distribution of bora ground orientations.

**Methodology**

To determine the orientations of bora ceremonial sites, we obtained data for 63 such sites from the following published literature: Mathews (1894, 1896a, 1896b, 1897a, 1907, 1917), Hopkins (1901), Towle (1942), Bartholomai and Breeden (1961), Anon. (1973), McBryde (1974), Steele (1984), Satterthwait and Heather (1987), and Bowdler (2001). We also obtained 1107 archaeological site cards from the NSW Aboriginal Heritage Information Management System (AHIMS) related to stone arrangements and ceremonial grounds.

We then filtered the data through a rigorous selection process, discarding data for any bora ground that failed to meet any of the following criteria:

1. The site is clearly described as a bora ceremonial ground;
2. The site is in NSW or southeast Qld (see Figure 3);
3. Measurements were made by an appropriately trained or qualified person (e.g. a surveyor or archaeologist);
4. The data are either first-hand, or second-hand from a trusted source;
5. There is unambiguous information on the direction from the large to the small circle; and,
6. a. Both rings and the pathway between them are identifiable, **or**
   b. Both rings and at least one opening are identifiable**, or**
   c. Only one ring is identifiable, but it has a clearly identifiable opening and there is unambiguous information as to whether it is the larger or the smaller ring.

The orientation of each site was measured from the centre of the largest circle to the centre of the smaller circle. If the second circle was missing, the orientation was taken from the centre of the circle to the middle of the opening. Measurements were either taken directly from the records given by the surveyor or measured from the survey with a protractor and ruler. We divided the azimuths into 16 bins (N, NNE, NE, ENE, E, ESE, SE, SSE, S, SSW, SW, WSW, W, WNW, NW, NNW), each with a width of 22.5º with N centred at 0°, NNE at 22.5°, and so on (see Table 1).

*Table 1:* *Azimuths of the inter-ordinal (left), ordinal (right top) and cardinal (right bottom) orientations (in degrees) used to bin the data in Figure 4, given in degrees. Each bin in the left-hand column has a width of 22.5° (±11.25°) centred on the Central Azimuth.*

| Orientation | Minimum Azimuth | Central Azimuth | Maximum Azimuth | Orientation | Minimum Azimuth | Central Azimuth | Maximum Azimuth |
|---|---|---|---|---|---|---|---|
| N | 348.76 | 0 | 11.25 | N | 337.5 | 0 | 22.5 |
| NNE | 11.26 | 22.5 | 33.75 | NE | 22.6 | 45 | 67.5 |
| NE | 33.76 | 45 | 56.25 | E | 67.6 | 90 | 112.5 |
| ENE | 56.26 | 67.5 | 78.75 | SE | 112.6 | 135 | 157.5 |
| E | 78.76 | 90 | 101.25 | S | 157.6 | 180 | 202.5 |
| ESE | 101.26 | 112.5 | 123.75 | SW | 202.6 | 225 | 247.5 |
| SE | 123.76 | 135 | 146.25 | W | 247.6 | 270 | 292.5 |
| SSE | 146.26 | 157.5 | 168.75 | NW | 292.6 | 315 | 337.5 |
| S | 168.76 | 180 | 191.25 | | | | |
| SSW | 191.26 | 202.5 | 213.75 | Orientation | Minimum Azimuth | Central Azimuth | Maximum Azimuth |
| SW | 213.76 | 225 | 236.25 | N | 316 | 0 | 45 |
| WSW | 236.26 | 247.5 | 258.75 | E | 46 | 90 | 135 |
| W | 258.76 | 270 | 281.25 | S | 136 | 180 | 225 |
| WNW | 281.26 | 292.5 | 303.75 | W | 226 | 270 | 315 |
| NW | 303.76 | 315 | 326.25 | | | | |
| NNW | 326.26 | 337.5 | 348.75 | | | | |

**Results and Analysis**

Of the 1170 sites obtained from the literature and AHIMS site card search, only 68 fit the specified selection criteria (Table 2, Figure 3). The 68 sites can be divided into two groups:

1) Those 46 sites for which the orientation has been recorded individually for each site; and,

2) Those 22 sites in Satterthwait and Heather (1987: Table 9), which provide the distribution of orientations without providing individual site orientations.

The 46 data points whose orientations are recorded individually, are provided in Table 2 and are grouped into 22.5° bins, shown as a histogram in Figure 4a. The histogram reveals a clear preference for the S, SW and W orientations. Table 2 contains 34 orientations in ordinal (N, S, E, W, NW, NE, SE, SW) directions, but only 12 orientations in inter-ordinal (NNE, ENE, etc.) directions. The low number of inter-ordinal orientations suggests either that the sites tend to be oriented on ordinal points, or that some authors rounded to the nearest ordinal point. To avoid a statistical bias in our results, we re-binned the data into the eight ordinal directions, dividing the counts of each inter-ordinal bin equally between the two neighbouring ordinal bins, resulting in eight 45° bins (Table 1, Figure 4b).

A preferred orientation to the south is evident (28% of the 46 data points), with lesser but significant preferences to the SW (17%) and W (15%) bins. The combined S, SW and W bins account for 61% of the total data points, while the remaining are evenly spread across the remaining bins. The highest peak of 12.5 orientations occurs in the S bin.

We compare this result with the 22 data points from Satterthwait and Heather (1987) (hereafter referred to as 'SH') by re-binning our data to four quadrants centred on the cardinal points (N, S, E, W), as shown in Figure 4c. This reveals a strong preference for the southern quadrant and a lesser but significant preference to the western quadrant, which, when combined, account for 74% of the 46 data points. The SH data (Figure 4d) is similar, with a significant preference for the southern quadrant (68% of the 22 data points), but with an even distribution among the remaining quadrants. We then combined our 46 data points with the 22 SH data points, resulting in Figure 4e. A clear preference for the southern quadrant is evident, with 35 of the 68 orientations (51%) falling in the S bin. This result is consistent with the Love hypothesis.

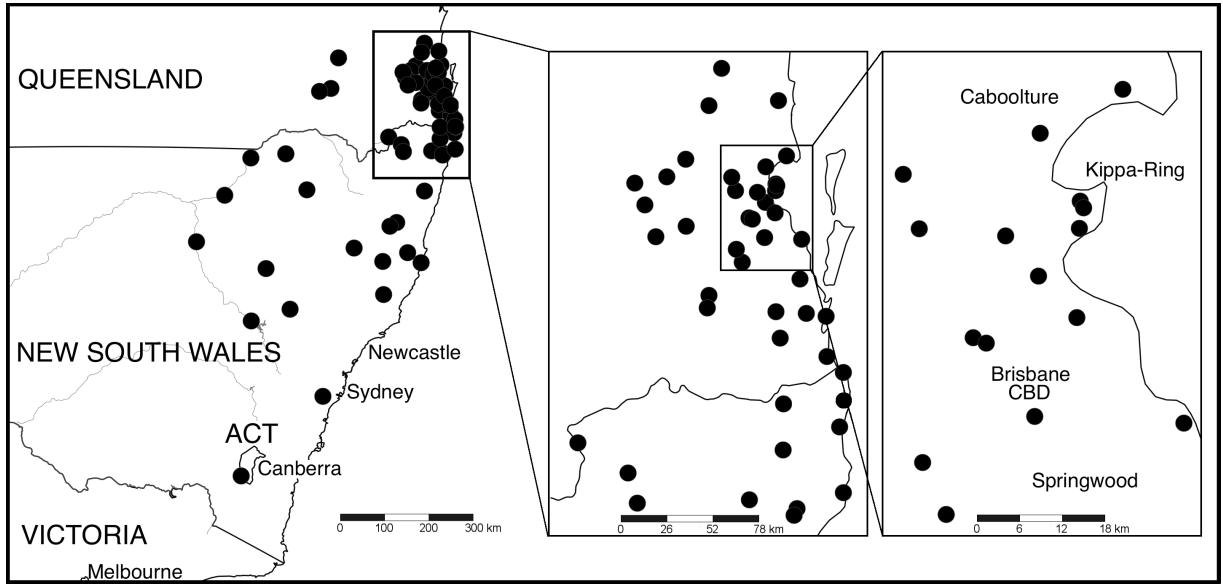

*Figure 3:* Locations of all bora grounds in Table 2, excluding the data from Satterthwait and Heather (1987). The coordinates given are adjusted to as to protect exact location of each site. The coordinates given are within 10 km of the site, which is why some appear over the sea.

To determine if this is a chance clumping of a random distribution of orientations, we conducted a Monte Carlo simulation, in which 68 orientations were distributed randomly in each of the bins shown in Figure 4e. We repeated this process 100 million times. In only 303 of the 100 million runs did the number in any one bin equal or exceed 35, from which we can conclude that the likelihood of the peak in Figure 4b occurring by chance is about $3 \times 10^{-6}$ or 0.0003%. We therefore conclude that this distribution is clearly not the result of chance, and that the constructors of the bora rings intentionally aligned most of them to the south quadrant.

*Table 2:* The bora grounds after filtering the original data through the selection criteria, grouped alphabetically by reference author, followed by the SH data. The coordinates given are approximate and do not reveal the exact location of the site. Data includes the site number, name and location. Also included is the site layout type: Type 1 consists of two circles and a pathway; Type 2 consists of two circles and one opening; Type 3 consists of one circle and an opening. All orientations are given in terms of cardinal, ordinal or inter-ordinal points measured from the large ring to the small ring, with the specific azimuth given where applicable. SH do not give coordinates or azimuths—only orientations in four quadrants (N, S, E and W).

| No. | Site Name | Location | Type | Orientation | Az | Source |
|---|---|---|---|---|---|---|
| 1 | Kogan | −27.2, 150.5 | 1 | W | 270 | Bartholomai and Breeden (1961) |
| 2 | Gurah | −29.4, 149.4 | 1 | E | 90 | Hopkins (1901) |
| 3 | Gurah 2 | −27.5, 153.3 | 2 | S | 180 | Steele (1984) |
| 4 | Wellington | −32.3, 148.6 | 2 | E | 90 | Mathews (1894) |
| 5 | Gundabloui | −29.1, 148.5 | 1 | WSW | 240 | Mathews (1894) |
| 6 | Wilpinjong Creek | −32.2, 149.5 | 1 | SW | 215 | Mathews (1894) |

| 7 | Eurie Eurie Run | −29.6, 148.1 | 1 | SW | 225 | Mathews (1894) |
|---|---|---|---|---|---|---|
| 8 | Bulgeraga Creek | −30.6, 147.3 | 1 | SSW | 202 | Mathews (1896a) |
| 9 | Camden | −34.3, 150.4 | 1 | NE | 45 | Mathews (1896b) |
| 10 | Murrumbidgee River | −35.5, 148.5 | 1 | S | 185 | Mathews (1897a) |
| 11 | Terry Hie Hie | −29.5, 150.9 | 1 | W | 270 | Mathews (1917) |
| 12 | Wyrallah | −28.5, 153.2 | 3 | SSW | 202 | McBryde (1974) |
| 13 | Brackenridge | −27.2, 153.1 | 3 | SE | 135 | McBryde (1974) |
| 14 | Weir River | −27.4, 150.4 | 1 | WNW | 298 | Mathews (1907) |
| 15 | Casino | −28.5, 153.2 | 3 | N | 0 | RRHS (1973) |
| 16 | Tucki Tucki | −28.6, 153.2 | 1 | SW | 225 | RRHS (1973), AHIMS 04-4-0024 |
| 17 | Lennox Head | −28.5, 153.4 | 3 | NNW | 315 | RRHS (1973) |
| 18 | Burleigh | −28.5, 153.3 | 1 | SSW | 202 | Steele (1984) |
| 19 | Nudgee | −27.2, 153.5 | 1 | W | 270 | Steele (1984) |
| 20 | Hilliard's Creek | −27.3, 153.2 | 2 | E | 90 | Steele (1984) |
| 21 | Tamborine | −27.5, 153.7 | 1 | S | 180 | Steele (1984) |
| 22 | Kippa Ring | −27.1, 153.5 | 1 | SSW | 202 | Steele (1984) |
| 23 | Sandy Creek | −26.5, 152.4 | 1 | S | 180 | Steele (1984) |
| 24 | Wooyung | −28.3, 153.3 | 1 | N | 0 | Bowdler (2001) |
| 25 | Kangaroo Flat | −31.1, 152.1 | 3 | SW | 225 | Bowdler (2001), AHIMS 03-5-001 |
| 26 | South Tweed Heads | −28.1, 153.3 | 3 | W | 270 | Bowdler (2001), AHIMS 04-2-009 |
| 27 | Bogangar | −28.2, 153.3 | 2 | SW | 225 | AHIMS 04-2-0133 |
| 28 | South Grafton | −29.4, 152.6 | 3 | SSE | 157 | AHIMS 12-6-0115 |
| 29 | Woolbrook | −30.6, 151.2 | 1 | NW | 315 | AHIMS 20-6-0022 |
| 30 | Diamond Flat/Petroi | −31.0, 152.4 | 1 | S | 180 | AHIMS 21-5-007 |
| 31 | Ruby Creek | −28.4, 152.1 | 1 | S | 180 | Towle (1942), AHIMS 03-5-0006 |
| 32 | Wheatley Creek | −28.5, 152.2 | 3 | WNW | 307 | AHIMS 03-5-0011 |
| 33 | Yellow Creek | −28.5, 152.2 | 1 | W | 270 | AHIMS 03-6-0028 |
| 34 | Nimbin Brookside | −28.4, 153.1 | 3 | W | 270 | AHIMS 04-4-0037 |
| 35 | Dyamberin Station | −30.2, 152.2 | 3 | NNE | 22 | AHIMS 21-2-0006 |
| 36 | Yooroonah | −30.3, 152.1 | 3 | NE | 45 | AHIMS 21-5-0012 |
| 37 | Timor Dam | −31.3, 149.2 | 2 | W | 270 | AHIMS 28-2-0002 |
| 38 | Richardson's Crossing | −31.1, 152.6 | 3 | N | 0 | AHIMS 30-3-0001 |
| 39 | Bundook | −31.5, 152.8 | 2 | S | 180 | AHIMS 30-5-0011 |
| 40 | Tyalgum | −28.2, 153.1 | 1 | SW | 225 | Steele (1984) |
| 41 | Alberton | −27.4, 153.2 | 1 | S | 180 | Steele (1984) |
| 42 | Amity Point | −27.5, 153.3 | 1 | E | 90 | Steele (1984) |
| 43 | Glenore Grove | −27.3, 152.2 | 3 | SSE | 157 | Steele (1984) |
| 44 | Milbong | −27.5, 152.4 | 1 | S | 180 | Steele (1984) |
| 45 | Cedar Creek | −27.9, 153.2 | 1 | SSE | 158 | Steele (1984) |
| 46 | Kunopia | −29.4, 149.4 | 1 | SW | 240 | Mathews (1896a) |
| 47 | Somerset Dam | | | 1 | | | Satterthwait and Heather (1987) |
| 48 | Samsonvale | | | 1 | | | Satterthwait and Heather (1987) |
| 49 | Samford | | | 1 | | | Satterthwait and Heather (1987) |
| 50 | Canungra | | | 1 | | | Satterthwait and Heather (1987) |
| 51 | Kippa Creek | | | 1 | | | Satterthwait and Heather (1987) |
| 52 | Mt Esk Pocket | | | 1 | | | Satterthwait and Heather (1987) |
| 53 | Camira | | | 1 | | | Satterthwait and Heather (1987) |
| 54 | Purga Creek | | | 1 | | | Satterthwait and Heather (1987) |
| 55 | Oakey Creek | | | 1 | | | Satterthwait and Heather (1987) |
| 56 | Waraba Creek | | | 1 | | | Satterthwait and Heather (1987) |
| 57 | Buaraba | | | 1 | | | Satterthwait and Heather (1987) |
| 58 | Dayboro West | | | 1 | | | Satterthwait and Heather (1987) |
| 59 | Walli Creek | | | 1 | | | Satterthwait and Heather (1987) |
| 60 | Upper Coomera River | | | 1 | | | Satterthwait and Heather (1987) |
| 61 | Moggill | | | 1 | | | Satterthwait and Heather (1987) |
| 62 | Toorbul Creek | | | 1 | | | Satterthwait and Heather (1987) |
| 63 | Keperra | | | 1 | | | Satterthwait and Heather (1987) |
| 64 | Petrie | | | 1 | | | Satterthwait and Heather (1987) |

| 65 | Lowood | 1 | Satterthwait and Heather (1987) |
| 66 | Kipper Creek | 1 | Satterthwait and Heather (1987) |
| 67 | Woolloongabba | 1 | Satterthwait and Heather (1987) |
| 68 | Jerribribillum | 1 | Satterthwait and Heather (1987) |

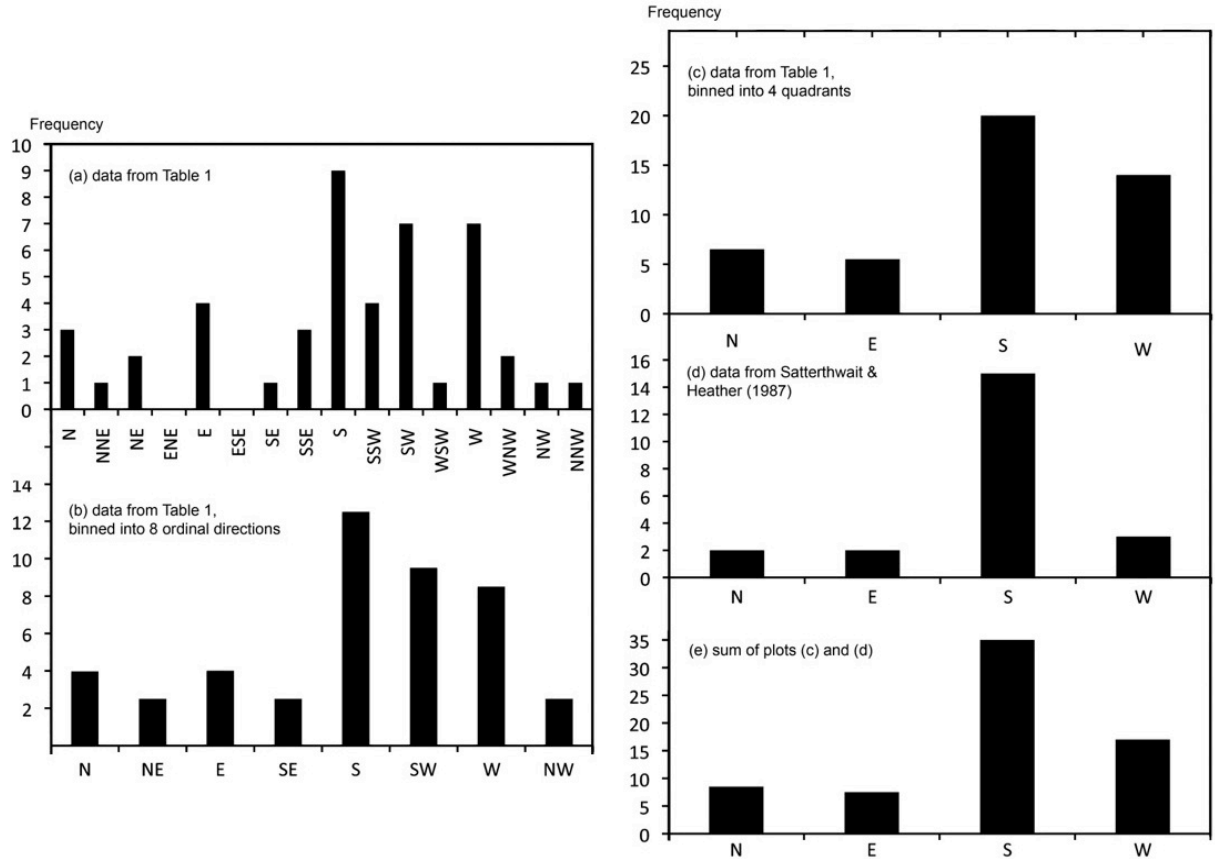

*Figure 4:* Orientations of bora sites using data from Table 1. Data in (a) is given in 22.5° bins, (b) is given in 45° bins, and (c) through (e) are given in 90° bins.

**Discussion and Conclusion**

We have shown that the bora grounds studied have a preferred orientation to southerly directions and that these orientations are not the result of chance, but were deliberate. The reason for this is not known, but it is consistent with the Love hypothesis that southeast Australian bora ceremonial grounds have a preferred orientation to the celestial emu in the Milky Way in the south-southwest skies. The celestial emu is in this position in the evening sky during the month of August, the time during which Winterbotham (1957) and Needham (1981) claimed that bora ceremonies were held. Hamacher et al. (2012) showed that linear stone arrangements in NSW also have a preferred orientation to the cardinal points, especially north-south orientations. Since many stone arrangements are ceremonial sites, this lends support to the claim that orientation is an important factor utilised by Aboriginal people when laying out ceremonial sites.

Although our analysis supports the Love hypothesis, it is not definitive evidence that bora grounds are oriented to the Sky Bora. Some researchers, including Winterbotham (1957) and Mathews (1894), state that many Bora ceremonies across NSW and Qld were held at various times throughout the year, which do not correspond to any particular orientation of the Milky Way. However, there is strong ethnographic evidence that the Milky Way is associated with the bora ceremony and we consider it likely that some ceremonies were timed, and bora sites oriented, such that the vertical Milky Way was visible above the path connecting the two circles during bora ceremonies. Additional research is necessary to understand these links; we are currently engaged in further research projects to explore this.

## Acknowledgements

We acknowledge the Aboriginal Elders and custodians, past and present, on whose land the bora sites are located. For archaeological data and literature sources, we thank staff from the State Library of NSW, the Australian Institute for Aboriginal and Torres Strait Islander Studies and the NSW Office of Environment and Heritage. This research made use of the TROVE and JSTOR databases, Google Maps, and the Stellarium astronomical software package. Hamacher conducted his component of this research while a PhD student at Macquarie University, but finished the analysis and writing as a staff member at the University of New South Wales.

## References

Anon. 1973 The layout of bora grounds. *Bulletin of the Richmond River Historical Society* 65:12–13.

Bartholomai, A. and S. Breeden 1961 Stone ceremonial grounds of the Aborigines in the Darling Downs area, Queensland. *Memoirs of the Queensland Museum* 13(6):231–237.

Berndt, R.M. 1974 *Aboriginal Religion*. Leiden: Brill Archive.

Black, L. 1944 *The Bora Ground: Being a Continuation of a Series on the Customs of the Aborigines of the Darling River Valley and of Central New South Wales, Part IV*. Sydney: F.H. Booth.

Bowdler, S. 2001 The management of Indigenous ceremonial ('bora') sites as components of cultural landscapes. In M. Cotter, W.E. Boyd and J. Gardiner (eds), *Heritage Landscapes: Understanding Place and Communities*, pp.1–19. Lismore: Southern Cross University Press.

Cairns, H.C. and B. Yidumduma Harney 2004 *Dark Sparklers*. Sydney: Hugh Cairns.

Collins, D. 1798 *An Account of the English Colony in New South Wales, Vol. I*. London: T. Cadell Jr and W. Davies.

Eliade, M. 1996 *Patterns in Comparative Religion*. Lincoln: University of Nebraska Press.

Frankel, D. 1982 Earth rings at Sunbury, Victoria. *Archaeology in Oceania* 17:83–89.

Fraser, J.F. 1883 The Aborigines of New South Wales. *Transactions and Proceedings of the Royal Society of New South Wales* 16:193–233.

Langevad, G. 1982 *Some Original Views Around Kilcoy. Book 1: The Aboriginal Perspective*. Queensland Ethnographic Transcripts 1. Brisbane: Archaeology Branch, Queensland Department of Aboriginal and Islanders Advancement.

Hamacher, D.W. 2012 *On the Astronomical Knowledge and Traditions of Aboriginal Australians*. PhD thesis by publication, Department of Indigenous Studies, Macquarie University. [*www.academia.edu/1905624/On_the_Astronomical_Knowledge_and_Traditions_of_Aboriginal_Australians*]

Hamacher, D.W., R.S. Fuller and R.P. Norris 2012 Orientations of linear stone arrangements in New South Wales. *Australian Archaeology* 75:46–54.

Hardie, B. 2004 *MRN9*. Aboriginal Heritage Information Management System, Card No. 51-6-0250. Sydney: Office of Environment and Heritage, NSW State Government.

Hartland, E.S. 1898 The 'High Gods' of Australia. *Folklore* 9(4):290–329.

Henderson, J. 1832 *Observations on the Colonies of New South Wales and Van Diemen's Land*. Calcutta: Baptist Mission Press.

Hopkins, A. 1901 Bora ceremony. *Science of Man* 4(4):62–63.

Howitt, A.W. 1884 The native tribes of south-eastern Australia. *Journal of the Anthropological Institute of Great Britain and Ireland* 13:432–459.

Howitt, A.W. 1904. *The Native Tribes of South-East Australia*. London: McMillan.

Hunter, J. 1793 *An Historical Journal of the Transactions at Port Jackson and Norfolk Island*. London: John Stockdale.

Jacob, T. 1991 *In the Beginning: A Perspective on Traditional Aboriginal Societies*. Perth: Western Australia Ministry of Education.

Johnson, D. 1998 *Night Skies of Aboriginal Australia: A Noctuary*. Oceania Monograph No. 47. Sydney: Sydney University Press.

Love, W.R.F. 1987 There is an emu on the bora ground. *Anthropological Society of Queensland Newsletter* 177:1–5.

Love, W.R.F. 1988 Aboriginal Ceremonies of South East Australia. Unpublished MA thesis, Department of Anthropology and Sociology, University of Queensland, Brisbane.

McBryde, I. 1974 *Aboriginal Prehistory in New England: An Archaeological Survey of Northeastern New South Wales*. Sydney: Sydney University Press.

Mathews, R.H. 1894 Aboriginal bora held at Gundabloui in 1894. *Journal of the Royal Society of New South Wales* 28:98–129.

Mathews, R.H. 1896a The bora, or initiation ceremonies of the Kamilaroi tribe. *Journal of the Anthropological Institute* 24:411–418.

Mathews, R.H. 1896b The burbung of the Wiradthuri tribes, Part 1. *Journal of the Royal Anthropological Institute* 25:295–318.

Mathews, R.H. 1897a The burbung of the Darkinung tribes. *Proceedings of the Royal Society of Victoria* 10(1):1–3.

Mathews, R.H. 1897b The burbung, or initiation ceremonies of the Murrumbidgee tribes. *Journal and Proceedings of the Royal Society of New South Wales* 31:111–153.

Mathews, R.H. 1907 *Notes on the Aborigines of New South Wales*. Sydney: Government Printer.

Mathews, R.H. 1910 Initiation ceremonies of some Queensland tribes. *Proceedings and Transactions of the Royal Geographical Society of Australia, Queensland* 25:103-118

Mathews, R.H. 1917 Description of two bora grounds of the Kamilaroi tribe. *Journal and Proceedings of the Royal Society of New South Wales* 51:427–30.

Needham, W. 1981 *Burragurra-A Study of the Aboriginal Sites in the Cessnock-Wollombi Region of the Hunter Valley*. Adamstown: Dobson and McEwan.
Norris, R.P. and D.W. Hamacher 2009 The astronomy of Aboriginal Australia. In D. Valls-Gabaud and A.

Boksenberg (eds), *The Role of Astronomy in Society and Culture*, pp.39–47. Cambridge: Cambridge University Press.

Norris, R. and P. Norris 2009 *Emu Dreaming: An Introduction to Australian Aboriginal Astronomy*. Sydney: Emu Dreaming Press.

Ridley, W. 1873 Australian languages and traditions. *The Journal of the Anthropological Institute Great Britain and Ireland* 2:257–275.

Roth, W.E. 1909 On Certain Initiation ceremonies. *North Queensland Ethnography* 12:166-85

Satterthwait, L. and A. Heather 1987 Determinants of earth circle site location in the Moreton region, southeast Queensland. *Queensland Archaeological Research* 4:1–33.

Smith, W. 1913 Sketcher, Aboriginal folk-lore. *The Queenslander*, 11 October, p. 8.

Stanbridge, W. 1861 Some Particulars of the General Characteristics, Astronomy, and Mythology of the Tribes in the Central Part of Victoria, Southern Australia. *Transactions of the Ethnological Society of London* 1:286-304.

Steele, J.G. 1984 *Aboriginal Pathways in Southeast Queensland and the Richmond River*. Brisbane: University of Queensland Press.

Towle, C.C. 1942 Bora ground near Ruby Creek, NSW. *Victorian Naturalist* 59(5):80–81.

Winterbotham, L. 1957 Gaiarbau's Story of the Jinibara Tribe of Southeast Queensland. Unpublished manuscript held on file at the Australian Institute for Aboriginal and Torres Strait Islander Studies, Canberra (MS #45/7460).

Wellard, G.E.P. 1983 *Bushlore - or this and that from here and there*. Perth: Artlook Books.